\documentclass[openacc=False]{rstransa}

\usepackage{subcaption}

\titlehead{Review}

\begin{document}

\title{Finding Black Holes: an Unconventional Multi-messenger}

\author{
Laura E. Uronen$^{1}$, Tian Li$^{2}$, Justin Janquart$^{3,4, 5, 6}$, Hemantakumar Phurailatpam$^{1}$, Sheung Chi Poon$^{1}$, Ewoud Wempe$^{7}$, Léon V. E. Koopmans$^{7}$, Otto A. Hannuksela$^{1}$}

\address{$^{1}$Department of Physics, The Chinese University of Hong Kong, Shatin, NT, Hong Kong\\
$^{2}$Institute of Cosmology and Gravitation, University of Portsmouth, Burnaby Rd, Portsmouth, PO1 3FX, UK\\
$^{3}$Nikhef – National Institute for Subatomic Physics, Science Park, 1098 XG Amsterdam, The Netherlands\\
$^{4}$Department of Physics, Utrecht University, Princetonplein 1, 3584 CC Utrecht, The Netherlands \\
${}^5$  Centre for Cosmology, Particle Physics and Phenomenology - CP3, Universit\'e Catholique de Louvain, Louvain-La-Neuve, B-1348, Belgium \\
${}^6$ Royal Observatory of Belgium, Avenue Circulaire, 3, 1180 Uccle, Belgium\\
${}^7$ Kapteyn Astronomical Institute, University of Groningen, P.O Box 800, 9700 AV Groningen, The Netherlands}

\subject{astrophysics, observational astronomy, cosmology, galaxies, relativity}


\corres{Laura E. Uronen\\
\email{laura.uronen@gmail.com}}

\begin{abstract}
A rather clear problem has remained in black hole physics: localizing black holes. One of the recent theoretical ways proposed to identify black hole mergers' hosts is through multi-messenger gravitational lensing: matching the properties of a lensed galactic host with those of a lensed gravitational wave. This paper reviews the most recent literature and introduces some of the ongoing work on the localization of binary black holes and their host galaxies through lensing of gravitational waves and their electromagnetically-bright hosts.
\end{abstract}

\maketitle



\section{Introduction}

Imagine a black hole: what do you see? Some of us will, perhaps, picture the striking visuals of the simulated supermassive Gargantua from the film \textit{Interstellar} (2014) or its real-life observational equivalent Messier 87 \cite{Messier87}. Others yet may imagine one of the LIGO-Virgo-Kagra (LVK) collaboration's animations showing two black holes whirling into one another against a bright galactic backdrop of stars. 



But of course, most of us will probably catch onto the rather obvious sticking point that will build the foundation of this paper and its science case: black holes are dark. The most apt picture when talking about stellar-mass black holes and binaries in the vast cosmic void that we observe in the LVK gravitational waves (GWs) is therefore simply a black hole on a black background. This does not make for very interesting visuals, but it is the central feature of the scientific challenges surrounding these common but extreme objects. So, is there a way to find them?

While traditional gravitational lensing of electromagnetic (EM) sources has been standing as an established pillar of GR observations for several decades already \cite{Dyson1920, Walsh1979, Shajib2024}, the detection of gravitational waves has opened the new avenue of gravitational lensing of GWs \cite{Ohanian1974, Degushi1986, Wang1996, Nakamura1998, Takahashi2003}. It has been proposed that lensing may have applications in binary black hole (BBH) localization as well \cite{Hannuksela2020}: if the GW from a binary is lensed, the EM-bright, galactic host of this binary will also be lensed. The simple theory is therefore: for a lensed gravitational wave, can we find a lens whose properties match those of the lensed GW? 

Typical multi-messenger astronomy has relied on multiple detections from the same source---such as the optical and GW identification of a kilonova~\cite{Abbott2017GW170817, Abbott2017MMpaper}. In this case, we take a rather more unconventional route: the multi-messenger signals do not originate from the same source. Instead, we consider the scenario where a BBH (emitting gravitational waves) is embedded in a galaxy (emitting light). The association of the two links the properties they exhibit: the lensed BBH and the lensed host galaxy must both intrinsically show lensing characteristics produced by the same lens.

The paper is structured as follows. In Section~\ref{sec:motivation}, we review BBH localization efforts so far and the unique benefits provided by lensing. Section~\ref{method} builds a concept of the path towards localizing a BBH, from lensed GW identification to sub-arcsecond localization. We address some of the challenges and applications of this method in Section~\ref{sec:limitations}, and draw this paper to a close with a summary in Section~\ref{sec:conclusion}.

\begin{figure}
    \centering
    \includegraphics[width=0.9\linewidth]{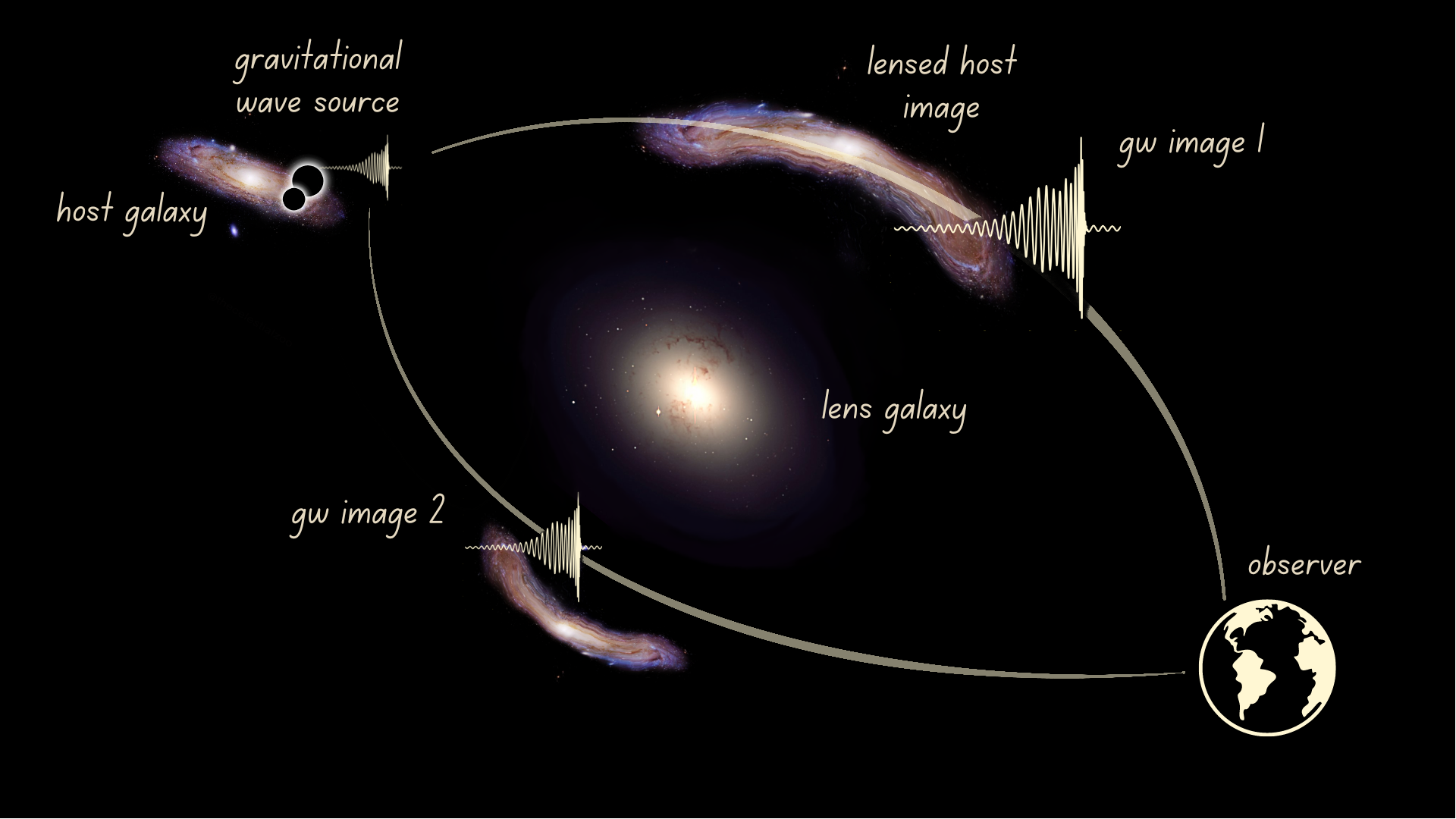}
    \caption{Schematic showing the idea behind this methodology for multi-messenger detections of lensed BBH: for a BBH embedded in its galaxy, if the gravitational wave is lensed, its host will also be lensed.}
    \label{fig:doodle}
\end{figure}

\section{Motivation}\label{sec:motivation}

\subsection{Black holes in single-messenger}

Some direct observations of stellar mass black holes have been made in the EM regime in the past. The most prominent systems are X-ray binaries, in which a stellar companion's material is stripped and falls into the black hole, as the material heats up and emits in the X-ray \cite{Casares2017}. Other observations rely on the lensing effect induced by a black hole and a star's light crossing paths, leading to observable microlensing \cite{Sahu2022}, or on radial velocity measurements in binaries \cite{Gu2019}, all relying on telescopes pointing at them to see them. But most of these black holes rest within our galaxy and leave us with the problem of understanding black holes throughout the Universe. 

Conversely, GWs have provided a uniquely direction-agnostic search for BBHs: gravitational waves are observed from any direction as long as the detectors are on. Additionally, these black holes lie at a variety of redshifts, though current detections still originate from the relatively nearby Universe reaching just past maximum redshifts of $z = 1$ \cite{Abbott2023GWTC3}. They also have the additional interesting feature of being generally heavier than the EM-observed counterparts (though, more recently in an exciting turn of events, evidence has been found of a black hole within our Milky Way falling for the first time within the higher mass ranges typically observed by the LVK \cite{Gaia2024}).

The problem with this direction-agnostic search is that retrieving the location of emission is harder. The LVK network's \cite{AdvancedLIGO, AdvancedLVK, Abbott2020, AdvancedVIRGO, KAGRA2020} localization of GW events relies on triangulation \cite{Fairhurst2018}, leading to a highly varying localization precision, largely limited by the number of detectors online \cite{Abbott2022GWTC21,Abbott2023GWTC3}. At best, as was the case of GW170814 at time of GW detection, are three-detector events that lead to sky localizations of roughly $\mathcal{O}(10)$ deg\textsuperscript{2} \cite{Abbott2017GW170814}. Most of the binaries detected by LVK are single- or two-detector events, with localizations $\mathcal{O}(100-10^4)$ deg\textsuperscript{2} \cite{Abbott2022GWTC21, Abbott2023GWTC3}. While the forecasted future sensitivities of aLIGO and aVirgo \cite{AdvancedLVK,Gupta2024}, detections by KAGRA \cite{KAGRA2020} and the addition of LIGO-India to the detector network \cite{ligoindiaproposal} may allow BNS sources to be consistently localized to within $\mathcal{O}(<10)$ deg\textsuperscript{2} for reasonable EM follow-up \cite{Fairhurst2014LIGOIndia}, the forecasted sky regions are still too large to lead to confident association with a host galaxy for a dark binary \cite{Petrov2022}. In this sense, at current predictions, the direct localization route is difficult for dark black hole binaries without any additional information. 

There are still interesting science cases to be done despite lack of precise localization. Statistical methods, such as dark siren cosmology, have been proposed as a means to measure the Hubble constant through the standard siren nature of GWs \cite{Muttoni2023}. The method is founded upon the idea that no specific host can be identified for the binary, but a statistical average would hold immense promise for cosmological studies. Some exciting prospects were outlined in the context of localization in terms of a few “golden” binaries that might lead to host localization based on gravitational-wave data alone~\cite{Chen2016}, but generally identifying a single host for any gravitational-wave event is considered challenging. 

\subsection{Multi-messenger lensing}\label{sec:mmlensing}
Here we consider a promising new avenue: multi-messenger lensing, though in a slightly different capacity than usual. As established, with lensing and GW acting as pillar consequences of GR, it is only natural that if a GW is lensed, then its host galaxy will also be lensed (Fig.~\ref{fig:doodle}). By searching for the host in the EM regime, it is then possible to match the parameters of the lensed GW to those of the EM lens \cite{Hannuksela2020,Wempe2022,Shan2023,Magare2023,Janquart2023_O3Search} (a more detailed description of the relationship between the lens and observed GWs is provided later on).

Because of the inherent dark nature of black holes, we cannot observe other messengers from these mergers directly; but this simultaneous lensing of the host and the GW can provide a new multi-messenger avenue to explore.

Current outlook for strong lensing detections remain promising for GWs from BBH \cite{haris2018identifying, ezquiaga2023identifying, goyal2021rapid,magare2024slick, Janquart2023_O3Search, Lo2023}, with up to 80\% of such detections being confidently identified at low false alarm probability~\cite{haris2018identifying} as long as time-delay windowing is incorporated into the analysis~\cite{Caliskan2023,Wierda2021}\footnote{The detection rates are somewhat lower for binary neutron star (BNS) or neutron star-black hole (NSBH) signals \cite{Buscicchio:2020bdq,Smith:2022vbp,Magare2023}, although the science-case made available from such detections would be immense, especially if multimessenger data were available.}. To date, none of the published events in the LVK data pass all the various strong lensing tests~\cite{Hannuksela2019,McIsaac2020,Dai2020,Liu2021,Abbott2021Lensing,Abbott2023Lensing,Janquart2023_O3Search}. However, the first GW lensing detection, when it arrives, will undoubtedly warrant follow-up opportunities, and it is necessary to develop the methodology to follow up on both dark and bright events. BNS lensing has a host of follow-up avenues, well-established by its unlensed predecessor GW170817 \cite{Abbott2017GW170817, Abbott2017MMpaper}, but this multi-messenger follow-up of BBH lensing can allow for all lensed GW events to be intriguing to a broad swathe of the astronomical community, be it for cosmology, population studies, tests of GR, or a variety of other applications.

The identification of a corresponding lens and host galaxy can also assist in GW lensing detections. Constraining the lens configuration can allow us to find lensed GW images that have remained hidden in the noise as sub-threshold events \cite{Li2023,Goyal2023}. With the lens system identified, we can do more targeted searches for these sub-threshold events \cite{Ng2024}. 

\subsection{Advantages}

We can summarize the benefits of combining both GW detections with EM observations.

From the gravitational wave side: 

\begin{itemize}
    \item[+] Time delay measurements: the gravitational wave measurement, due to the LVK observatories' submillisecond precision timing measurements, can be obtained to $\mathcal{O}(<1)$ ms. \cite{Sullivan2023}
    \item[+] GW waveform is determined precisely by GR.
    \item[+] Gravitational waves are less susceptible to stellar-mass microlensing \cite{Cheung2021}.
    \item[-] The magnification is degenerate with the luminosity distance: the lensed gravitational wave is no longer a standard siren, and we instead measure the effective, or apparent, luminosity distance. 
\end{itemize}

From the EM observations, we can gain additional features which solve degeneracies in the GW regime, such that these two regimes complement one another:

\begin{enumerate}
    \item[+] Detailed lens reconstruction can be obtained from a high-resolution image.
    \item[+] Source redshift can be constrained to break the distance-magnification degeneracy.
\end{enumerate}

While this article will focus mainly on lensed BBHs due to the generally-intrinsic lack of EM counterpart, this approach is generic to any lensed GW without EM counterpart. There are scenarios where BNS are too faint to be observed in the EM, and it is currently unknown whether NSBH give an observable EM signal \cite{Zhu2021}. This method can then be used for any GW source. But as BBHs form the bulk of the detections made by the LVK, we will be mainly referring to these throughout this text.

\section{Road map to a black hole}\label{method}

To illustrate the process of lensed BBH localization, we draw up a simple roadmap to describe the approximate process which, when a lensed GW is identified, we can use to theoretically locate the binary. This road map will begin at the stage where we identify individual GW events, and tests the boundary of how far it may be possible to go in narrowing down the location of the BBH.

It must be noted that this method works only for multiple detected images: if we identify GW lensing for a single image with no other GW counterparts, as there are no possible measurements of relative time delays or magnifications we cannot use this method for lens-matching. 

\begin{figure*}

    \begin{subfigure}[t]{0.5\textwidth}

        \centering
        \includegraphics[width=\linewidth]{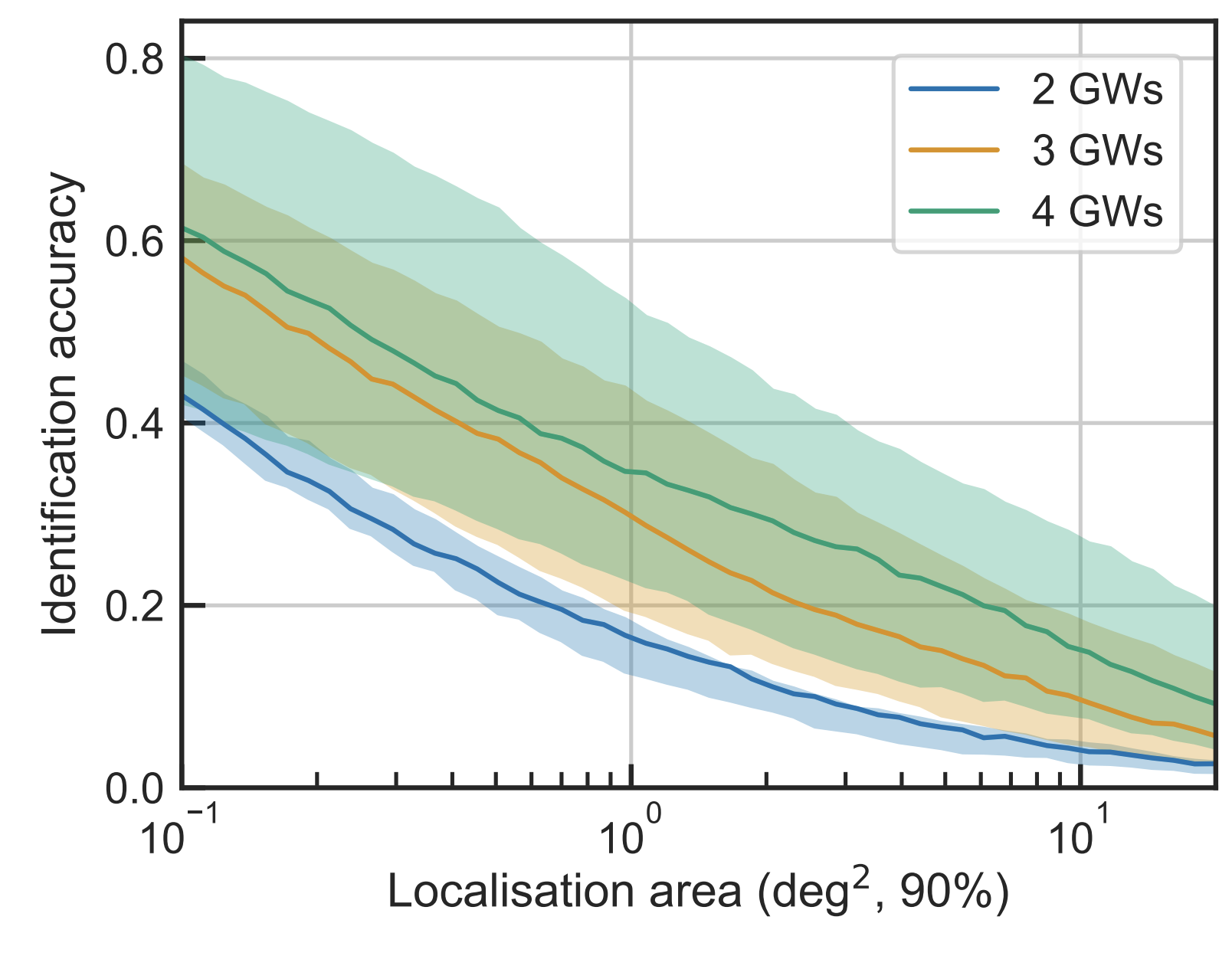}
        \caption{}
        \label{fig:WempeSkyloc}
    \end{subfigure}
    \begin{subfigure}[t]{0.5\textwidth}

        \centering
        \includegraphics[width=\linewidth]{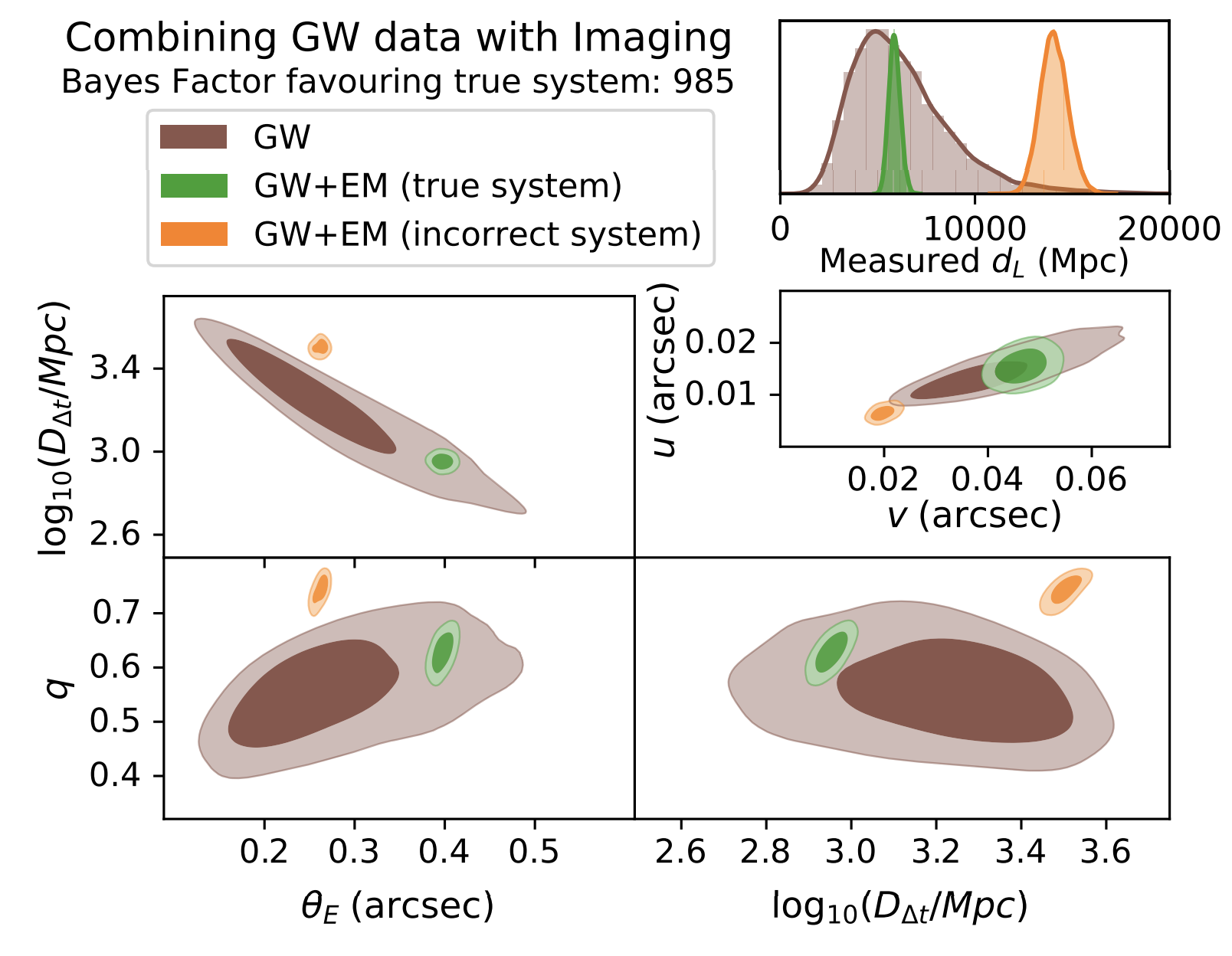}
        \caption{}
        \label{fig:WempeWrongSys}
    \end{subfigure}
    
    \caption{Figures from \cite{Wempe2022}. (a) For \textit{Euclid}-like imaging, the host identification accuracy mainly depends on the sky localization. (b) Lens parameter matching between a lensed GW event and two EM host candidate lenses for \textit{Euclid}-like imaging. The joint lensing parameter analysis with the wrong system (orange) shows a significant mismatch with the GW-only analysis, while the correct system (green) falls within the distributions from GW alone.}
\end{figure*}

\subsection{LVK detections}

We assume that a lensed gravitational wave has been identified. Initially, the sky localizations of the individual images will be anywhere from $\mathcal{O}(10-10^4)$ deg\textsuperscript{2}, as seen in the typical skymaps provided for different events by the LVK collaboration, depending on the number of detectors online at the time of image detection \cite{Abbott2022GWTC21, Abbott2023GWTC3}.

To determine if these images are lensed counterparts of one another, they are jointly analyzed through parameter estimation---a process similar to EM lens reconstruction (for descriptions of tools that do this, see Ref.~\cite{Janquart2022GOLUM} and Ref.~\cite{Lo2021}). This joint analysis will consider all the parameters of the different lensed images, and find their joint highest-likelihood values. For sky localization, this leads to a significantly narrowed-down localization posterior. The exact localization will depend greatly on the number of images identified (another caveat explored further later on), but the subsequent joint sky localization can be of the order of $\mathcal{O}$(1-10) deg\textsuperscript{2} \cite{Janquart2021, Lo2023Hanabi}. For events in which the initial sky location could be several orders of magnitude more ill-defined, this is a great improvement and yields a final sky location that is far more reasonable to search through.

\subsection{Lens association}

Having identified the smaller sky region within which the host is likely to be, we turn to search for the EM host.

\begin{figure*}

    \begin{subfigure}[t]{0.5\textwidth}

        \centering
        \includegraphics[width=0.85\linewidth]{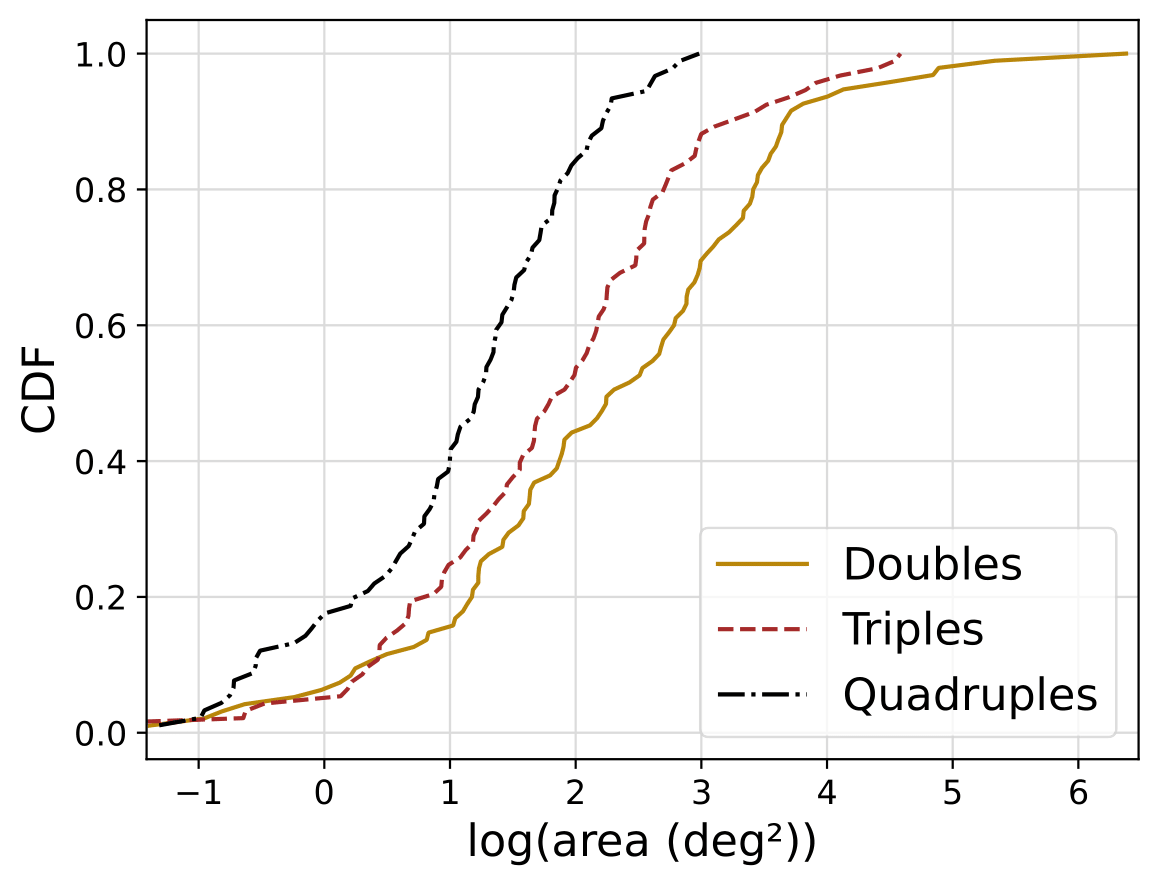}
        \caption{}
        \label{fig:SkyHistplot}
    \end{subfigure}
    \begin{subfigure}[t]{0.5\textwidth}

        \centering
        \includegraphics[width=\linewidth]{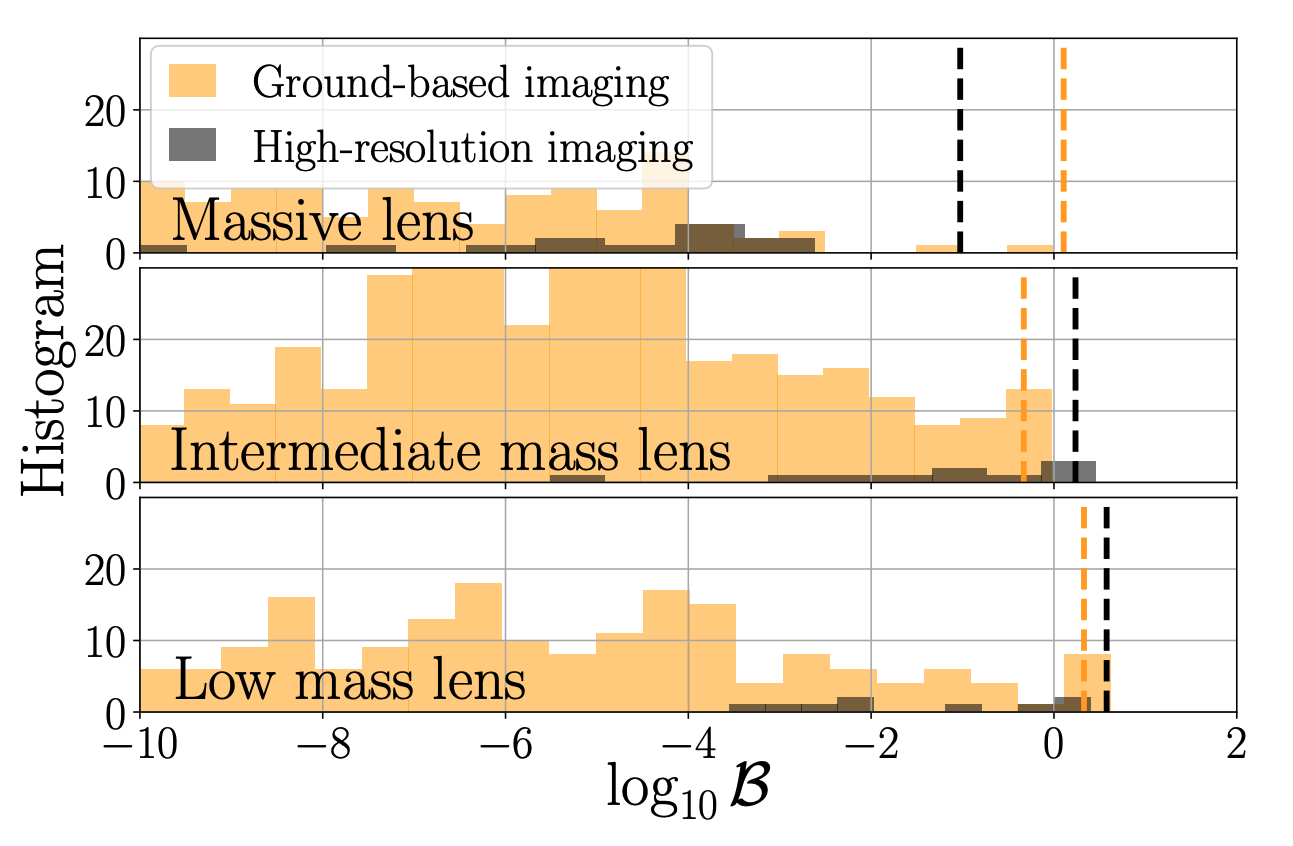}
        \caption{}
        \label{fig:HannukselaBayes}
    \end{subfigure}
    
    \caption{
    (a) Cumulative histogram showing the sky localization distribution for different numbers of images. As the number of images increases, the sky region distribution decreases as expected. The feature at small areas is due to low number statistics, and the fact that the datasets for different image numbers are independent systems (as opposed to re-analyses of the same systems with different image numbers). (b) Figure from \cite{Hannuksela2020}. A lensed GW is injected in three lens configurations. For each, 550 lenses in the sky region are reconstructed with ground-based imaging (yellow) to get the Bayes factor of association for each lens. Then the 11 highest-ranked lenses are reconstructed with high-resolution follow-up (black). The true host's Bayes factor is shown in each case with the dashed lines. For the high mass case, the host is unambiguously identified, while for the others the candidates are narrowed down to a handful.}
\end{figure*}

A benefit of this approach is that the EM follow-up need not be immediate. Since we are searching for a lensed galaxy, there is little risk of the observation fading. In fact, archival searches through lens catalogues, such as Ref. \cite{Vujeva2024}, are also a plausible way to search for the host, allowing for less time-sensitive searches to be completed. An example of such a search is done in Ref.~\cite{Janquart2023_O3Search}, and Fig.~\ref{fig:WempeSkyloc} shows the importance of the localization region size to the identification accuracy. The lenses identified in this sky region can be modelled together with the GW as shown in Fig.~\ref{fig:WempeWrongSys}, and ranked according to the likelihood with which they are associated with that particular lensed GW event. The link between image number and sky localization is also shown in Fig.~\ref{fig:SkyHistplot}, showing different simulated lensed systems with different numbers of detected images and their joint sky localizations. The association can be tested through Bayesian inference, allowing for the lenses to then be ranked according to their Bayes factor, effectively acting as a ranking statistic, which we will describe in more detail in the next section.

Naturally, the more common the lens in question is, the more there will be alternative lenses within the region being searched for. However, as shown in Fig.~\ref{fig:HannukselaBayes}, in the case of particularly large or exotic lenses, this will prove more effective due to a more limited number of corresponding candidate lenses. 

\subsubsection{A mathematical interlude}

Here we take a brief detour to cover the fundamental basics of this association. We start with the parameters that translate the effect that lensing has on GW, and proceed to the Bayesian association test. 

Under strong lensing, three effects are introduced in the wave equation in comparison to the unlensed waveform \cite{Dai2017}: 
\begin{align}
    h_L(f) &= \sqrt{\mu} e^{-2\pi i(f\Delta t - \frac{n}{2})} \tilde{h}_U(f)\,.
\end{align}
The changes to the waveform come from the magnification $\mu$, a time delay $\Delta t$, and a phase factor $n$. The latter is related to the image parity, and while it will be generally ignored for the purposes of this paper, it can play an important role in identifying the correct lens for the case of rarer image parities \cite{Keeton2003}.

These properties, in particular the magnification and time delay, also directly arise from the lens equation~\cite{Schneider1992}: 
\begin{align}
    \Vec{\beta} = \Vec{\theta} - \Vec{\alpha}(\Vec{\theta} )\,, 
\end{align}
where $\Vec{\theta}$ is the source position and $\Vec{\beta}$ is the image position, and $\Vec{\alpha}$ is the reduced deflection angle which relates to the lens model used.

The magnification and the time delay are given by~\cite{Wambsganss1998}:
\begin{align}
    \mu = \frac{\Vec{\theta}}{\Vec{\beta}} \frac{d\Vec{\theta}}{d\Vec{\beta}}\,, 
\end{align}
\begin{align}
    \Delta t = \frac{1 + z_l}{c} \frac{D_s D_l}{D_{ls}} \left( \frac{1}{2}(\Vec{\theta} - \Vec{\beta})^2 - \psi(\Vec{\theta}) \right)\,, 
\end{align}
where $\psi(\theta)$ is the gravitational potential at the source position, $D_l, D_s, D_{ls}$ respectively represent the angular diameter distances to the lens, to the source, and between the source and lens, and $z_l$ the lens redshift.

We present here a formal Bayesian derivation to test the association between a candidate lens and a (confirmed) lensed GW event. This is not vital to understand the principle of the association test, but it remains important in establishing the method through which we actually quantify this association. The formalism is based on Refs.~\cite{Hannuksela2020, Wempe2022} which established a similar association test.

We start from the odds ratio, $\mathcal{O_N^A}$, of the association hypothesis $\mathcal{H_A}$ against the null hypothesis $\mathcal{H_N}$:
\begin{align}
  \mathcal{O^A_N} = \frac{p(\mathcal{H_A} | d_{GW}, d_{EM})}{p(\mathcal{H_N} | d_{GW}, d_{EM})} = \frac{p(\mathcal{H_A})}{p(\mathcal{H_N})} \frac{p(d_{GW}, d_{EM} | \mathcal{H_A})}{p(d_{GW}, d_{EM} | \mathcal{H_N})} = \mathcal{P^A_N}\mathcal{B^A_N}\,,
\end{align}
where $\mathcal{P_N^A}$ refers to the prior odds, and $\mathcal{B_N^A}$ to the Bayes factor. We assume, for a fairly complete catalogue, that 
$$\mathcal{P^A_N} = 1/N_{lenses}\,.$$ 
That is, no lens is a priori likelier than any other lens in the sky region,\footnote{Note that a more massive stellar content in the galaxy will increase the likelihood that the galaxy hosts the black hole, but we neglect this here for simplicity, although it is discussed in Ref.~\cite{Wempe2022}.} but the more lenses there are in the identified localization before this association test, the less likely it is that we have found the correct lens with a single test.\footnote{For incomplete catalogs, we cannot know the exact number of lenses in the sky region, in which case the prior odds must be assumed to largely disfavor association and might be estimated based on the electromagnetic lensing estimates instead of proportional to number of lenses.} Identification is then primarily a question of finding a strong enough association to ‘beat’ the prior odds. We therefore focus on the Bayes factor.

The Bayes factor favoring association with a particular lens system is:
\begin{align}
    \mathcal{B^A_N} &= \frac{p(d_{GW}, d_{EM} | \mathcal{H_A})} {p(d_{GW} | \mathcal{H_N})p(d_{EM} | \mathcal{H_N})}= \frac{\mathcal{Z_A}}{\mathcal{Z}_{GW}\mathcal{Z}_{EM}} \,,
\end{align}
where $\mathcal{Z}_A$ is the evidence in favor of the system being associated with the GW. We will henceforth drop the hypothesis term $\mathcal{H}_X$ for readability and use a subscript to mark the corresponding hypothesis (e.g., we use $p_A(d_{GW},d_{EM})$ in place of $p(d_{GW},d_{EM}|\mathcal{H_A})$).

With gravitational waves, we can measure all the binary black hole detector-frame parameters $\Vec{\theta}_{BBH}^\prime$ except from so-called effective luminosity distances of each image $\Vec{D}_L^{eff}$, the time delay $\Delta \Vec{t}$ and the Morse phase $\Delta \Vec{n}$ 
(i.e., $\Vec{\theta}_{GW}^\text{all} = \{\Vec{\theta}_{BBH}^\prime, \Vec{D}_L^{eff}, \Delta \Vec{t}, \Delta \Vec{n}\}$). Since only the effective luminosity distances, arrival times, and the Morse phases are affected by strong lensing, we can marginalise over the other parameters to reduce the set of parameters we examine from the GW side (i.e., we focus on $\Vec{\theta}_{GW}^\text{reduced} = (\Vec{D}_L^{eff}, \Delta \Vec{t}, \Delta \Vec{n})$ parameters in place of $\Vec{\theta}_{GW}^\text{all}$) when performing joint lens modelling. From the EM side, one can obtain the lens parameters from lens reconstruction $\Vec{\theta}_{lens}$, the lens redshift $z_l$, and the source redshift $z_s$ (as an example of lens modelling, see \textsc{lenstronomy} \cite{Birrer2018}). That is, $\Vec{\theta}_{EM} = \{\Vec{\beta}, \Vec{\theta}_{lens}, z_l, z_s\}$. A joint analysis of the two allows for better lens modelling and enables one to retrieve the compact binary merger's angular position in the source plane $\Vec{\beta}$.

\begin{figure*}
    \begin{subfigure}[t]{0.33\textwidth}

        \centering
        \includegraphics[width=\linewidth]{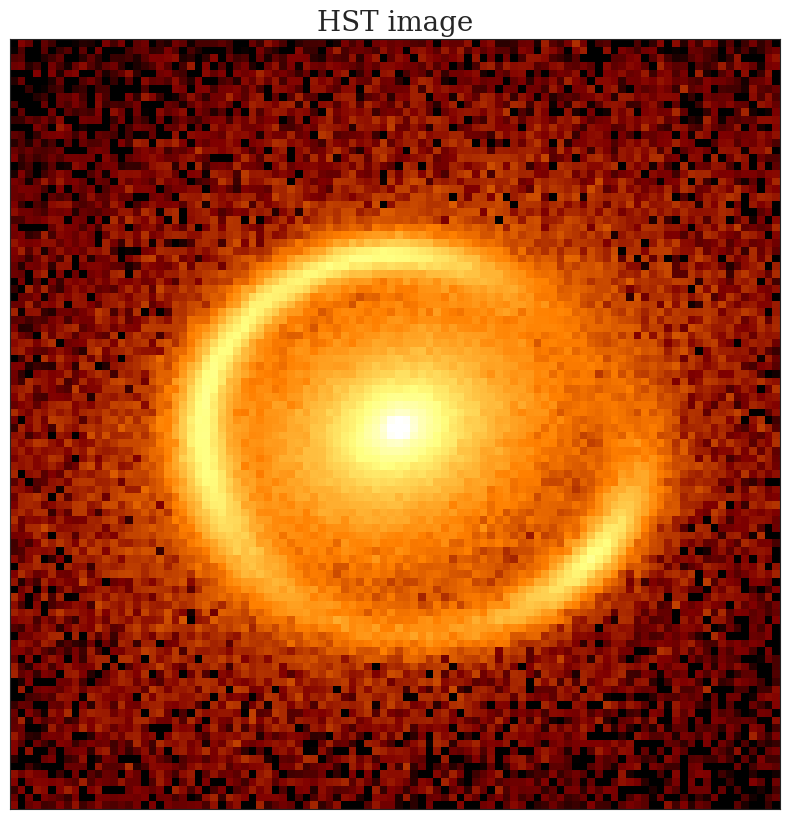}
        \caption{}
        \label{fig:hst}
    \end{subfigure}
    \begin{subfigure}[t]{0.33\textwidth}

        \centering
        \includegraphics[width=0.985\linewidth]{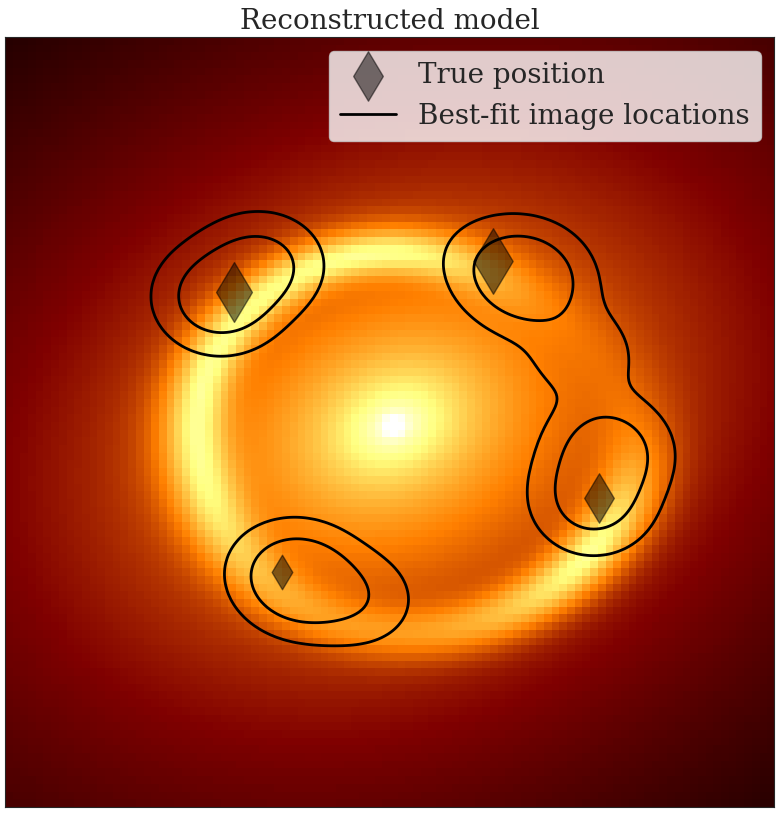}
        \caption{}
        \label{fig:imageplane}
    \end{subfigure}
    \begin{subfigure}[t]{0.33\textwidth}

        \centering
        \includegraphics[width=\linewidth]{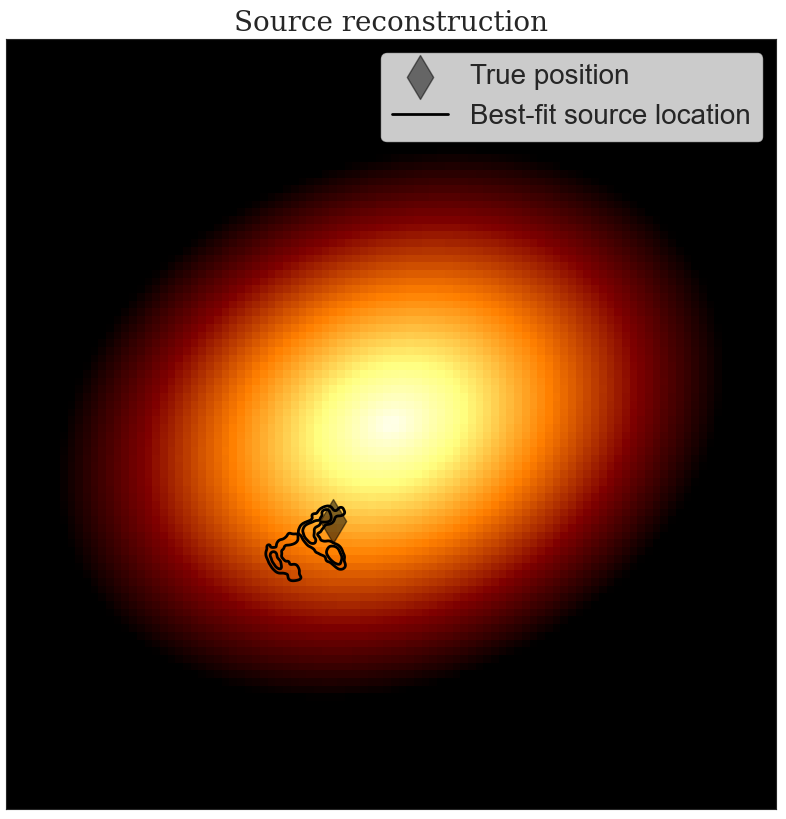}
        \caption{}
        \label{fig:sourceplane}
    \end{subfigure} 
    
    \caption{Demonstration of the joint reconstruction of a simulated lens and GW system and pin-point localization, with 1-$\sigma$ and 2-$\sigma$ outlines. (a) Simulated large galaxy lens system at HST resolution. (b) EM reconstruction of the image plane with reconstructed GW image positions and true GW image positions. (c) EM reconstruction of the source plane with reconstructed GW source position and true GW source position. As can be seen, the source and image positions are well-retrieved with some degeneracy due to lens modelling uncertainties.}
    \label{fig:simulation}
\end{figure*}

The GW lensing parameters are direct functions of the EM lensing parameters, as shown before, so we can marginalize over the GW lensing parameters fully and reduce the association hypothesis only to the set of parameters obtained from the EM observations. It can be shown (derivation in Appendix~A) that the evidence is given by:
\begin{align}
    \mathcal{Z_A} = p_\mathcal{A}(d_{EM}) 
     \int{p_\mathcal{A}(d_{GW} | \Vec{\theta}_{lens}, z_s, z_l, \Vec{\beta}}) p_\mathcal{A}(\Vec{\beta}) p_\mathcal{A}(z_s) p_\mathcal{A}(z_l) p_\mathcal{A}(\Vec{\theta}_{lens} | d_{EM})  d\Vec{\theta}_{EM} d\Vec{\beta}\,.
\end{align}
We take the lens redshift to be a delta function $p(z_l) = \delta (z_l - z_{l, measured})$ (typically, these can be measured for known lenses to a satisfactorily high precision with spectroscopy). If the source redshift is also known, this can be similarly marginalized out, but assuming that we do not know the source redshift this is kept in the calculation (this may be the case in archival data, prior to a dedicated follow-up \cite{Langeroodi2023}). 
By cancelling the EM evidences, the Bayes factor becomes:
\begin{align}
    \mathcal{B^A_N} = \frac{\int{p_\mathcal{A}(d_{GW} | \Vec{\theta}_{lens}, z_s, \Vec{\beta}}) p_\mathcal{A}(\Vec{\beta}) p_\mathcal{A}(z_s) p_\mathcal{A}(\Vec{\theta}_{lens} | d_{EM})  d\Vec{\theta}_{EM} d\Vec{\beta}}{p_\mathcal{N}(d_{GW})}\,.
\end{align} 
It is rather difficult to obtain the likelihood $p_\mathcal{A}(d_{GW} | \Vec{\theta}_{lens}, z_s, \Vec{\beta})$ directly. However, as these lens parameters yield the gravitational-wave observables, we can express the gravitational-wave observables as a function of these parameters and obtain the likelihood as a function of the relative time delay between two images $\Delta t_{ij}$ (since we cannot measure the arrival time independently) and effective luminosity distance $D_L^{eff}$. We can use Bayes' theorem and cancel the GW evidences to obtain:
\begin{align}
    \mathcal{B_N^A} &=  \int{\frac{p_\mathcal{A}(\Delta t_{ij}(\Vec{\theta}_{EM}), D_L^{eff}(\Vec{\theta}_{EM}) | d_{GW})}{p_\mathcal{A}(\Delta t_{ij})p_\mathcal{A}( D_L^{eff})}p_\mathcal{A}(\Vec{\beta}) p_\mathcal{A}(z_s) p_\mathcal{A}(\Vec{\theta}_{lens} | d_{EM})  d\Vec{\theta}_{EM} d\Vec{\beta}}\,.
\end{align}

\subsubsection{Dark siren reconstruction}

In addition to a direct search through all the lenses in the sky region, it is far more useful---and gentler on the precious computational resources available---to narrow down the list of candidate lenses first before rushing to reconstruct all of them. 

From the GW posteriors, it is possible to obtain some sense of the lens parameters. GWs provide limited information about the lens, unlike a complete EM picture of the lens does. As there are only three parameters introduced into the GW waveform from lensing, a great deal of degeneracies will exist in trying to reconstruct the lens characteristics. However, Ref. \cite{Poon2024} showed that a subset of the lens parameters can be obtained to represent the lens at the origin of the observed phenomenon. 

For a given sky region and identified lenses within it, it is then possible to obtain an estimate of the parameter space within which the most likely lenses are (Ng \textit{et al., in prep.}). From there, a shortlist of lenses can be used to determine the lenses that will need reconstruction, thus sparing a great deal of time and effort being wasted on unlikely lenses.

\subsection{Pin-point localization}

We now make a few additional assumptions: that we have observed the host and that we have also successfully identified the host galaxy of the GW. How far can we push the envelope? 

We can proceed to test how precisely we can localize the BBH with the same likelihood as the association. If the lens can produce the time delays exhibited by the GW, it will also mean that there is a location in the parameter space of the lens and the source plane from which this GW originated. We can then assume that if the lens produced the GW, we can retrieve the location in the source plane for this GW, and thus finding the sub-arcsecond localization of the GW. This joint reconstruction for a simulated system is shown in Fig.~\ref{fig:simulation}, in which the reconstruction for GWs was done with \textsc{GOLUM} \cite{Janquart2022GOLUM} and the EM reconstruction was completed using \textsc{lenstronomy} \cite{Birrer2018}.

We can expect that this will not be exact: since the most common lens models in use have some elements of symmetry, there will be some degeneracy in the localization of this GW, as well as lens modelling uncertainties and inherent to lensing. It is also possible that substructures can add some uncertainty to the time-delay modelling with time delay anomalies \cite{Liao2018}, similar to flux-ratio anomalies \cite{Mao1998}---however, we already account for this in increasing the time-delay uncertainties relative to lens modelling uncertainties, rather than keeping the time-delay uncertainty at the LVK precision.
Ultimately, this can be reduced by high-resolution imaging and kinematic data of the lens (to resolve the mass-sheet degeneracy \cite{Falco1985,Saha2000, Schneider1992}), detailed lens modeling and more complex mass models. Regardless, if the lens has a reasonable chance of producing the observed GW, we can retrieve the smaller sub-region of the source plane from which the signal came. 

\section{Future of black hole multi-messenger}\label{sec:limitations}

\subsection{Challenges}

\subsubsection{LVK lensing detections}

The first challenge to address should, naturally, be the challenge of identifying strong lensing. However, a variety of tests and pipelines exist to identify GW lensing~\cite{Dai2020, Ezquiaga2021, Goyal2023, haris2018identifying, Janquart2021_HOM,Janquart2021, Janquart2022GOLUM, Janquart2023Golum, Li2023, Liu2021, Liu2023, Lo2021, Lo2023, McIsaac2020, Mishra2023, More2022, Vijaykumar2022, Wang2021, Wierda2021, Wright2022}, and we expect the identification of GW lensing to be a matter of time (for a more complete review of GW lensing pipelines, see Ref.~\cite{Hannuksela2019,Abbott2021Lensing, Abbott2023Lensing, Janquart2023_O3Search}).

Following the identification of lensing, the challenge that arises is the GW sky localization. The sky localization from the joint analysis decreases with the number of detectors online, as it does with the number of detected images. Ref. \cite{Wempe2022} showed that the lens identification accuracy is highly dependent on the sky localization.

The good news is that we expect this problem to continuously lessen in importance: with detection forecasts hopeful from KAGRA and LIGO-India, the increased sensitivities in the LIGO and Virgo detectors at upcoming stages as well as the advent of next-generation detectors (such as Cosmic Explorer \cite{CE2017, CE2019, CE2021} and the Einstein Telescope \cite{ET2010, ET2011}), we expect to observe more super-threshold GW detections, as well as reducing the sky localization at primary observation \cite{AdvancedLVKloc, Li2022}. Hence, while this problem persists for the moment, we will see its importance drop as GW and GW lensing detections improve. 

We also do not include cluster lensing at this stage. In theory, should a generic cluster lens model be created, this method can be extended to testing associations between lensed GW events and cluster lenses. However, the problem currently remains that no such lens model exists apart from the use of analytical models as approximations~\cite{Meneghetti2003}\footnote{This can be done on a case-by-case basis for clusters where detailed lens modeling does exist (e.g. from lensing maps created by different collaborations available through the MAST archive  at \url{https://archive.stsci.edu/index.html}). Unless the cluster has been studied, approximations are unlikely to allow us to draw conclusions about clusters without such models.}. We therefore cannot draw any conclusions about cluster lensing, but with the developments being made towards cluster lens modelling, it will likely be a question of time and computational power before cluster lens reconstructions can be completed to a high enough degree of precision to be included trivially into the association test. 

\subsubsection{Finding the correct lens}

The next challenge comes from the EM side. We currently stand with catalogs summing to some $\mathcal{O}(10^3)$ known galaxy lenses \cite{Faure2008, Anguita2012, Jacobs2019, Li2021, Gaia2024Known}, with the most recent substantial addition coming from the Dark Energy Survey which at some estimates doubled the number of known galaxy lenses through machine learning methods of lens finding \cite{Jacobs2019}. This is expected to be only a fraction of the number of galaxy lenses that exist, facing a number of problems such as too small Einstein radii at current resolutions, or too faint to be observed by existing telescopes, or simply existing in regions that have not been surveyed to sufficient depth to identify lenses. This catalogue incompleteness means that we cannot presume, at this time, to find the correct lens without a dedicated follow-up---while the expected redshifts of current GW sources could mean that it is plausible for the lens to have already been observed, it is not an assumption that can safely be made. Hence, without a large enough Bayes factor of association to beat the prior odds (that must be assumed largely disfavoring association until we know to better certainty the number of lenses in a given sky region), we cannot claim to have found the host, even with a decent Bayes factor. This problem can be relieved in two ways: first with better-localized events, the number of lenses in the sky region will naturally drop, and second with catalog completeness through the advent of future telescopes like \textit{Euclid} \cite{Euclid2022}, LSST \cite{LSST2022} and \textit{Roman Space Telescope} \cite{RST2015}, which are expected to increase the number of lenses found to $\mathcal{O}(10^5)$ and increase our confidence in the association test \cite{Collett2015}. 

\textbf{Error 404: host not found?} One feature that has been observed with lensed supernovae has been that for a number of them, the identified host is far from the supernova itself \cite{Zinn2011}. The supernova, in some cases, either originates from a host too faint to detect (a problem related to the previous one, amended by future instruments) or otherwise from a host simply too far away from the lensed point source \cite{Pierel2023}. The latter case may be indicative of high-velocity kicks, and whether black holes receive these, and indeed the kick velocities themselves, remain a matter of some contention between simulations, GWs, and EM observations \cite{Callister2021, Varma2022}. It is therefore possible that in certain cases a lensed host simply cannot be found. If we were to assume that we expect most lenses in the sky region are known for a particular lensed GW, but no strong association can be made against the prior odds, it would imply that some lenses have been missed and the host cannot be found. In turn, non-detection of association between merger and host can allow us to place limits on host redshifts as well as kick velocities at BBH inception, a direction we are also looking to pursue in the future. 

\subsubsection{Lens imaging}

Aside from the lens finding, the EM lens reconstruction is a limiting factor. Assuming that we have a lensed gravitational-wave, and the host is observed, we can then proceed to reconstruct the lenses in the sky region (or at least the highest-priority ones). In this case, we may run into a new problem: the image resolution. For smaller, more distant sources, or lower-resolution survey instruments, these reconstructions will inevitably carry with them large modelling errors, degrading the results of the association test.

A neat and simple solution would of course be ground-based, high-resolution follow-up. For this, telescopes such as the Extremely Large Telescope (ELT) \cite{Padovani2023} make ideal candidates. With high-resolution telescope time being sought-after, these will be limited to either cases where high-resolution archival data exists already, or cases where follow-up is warranted.

\subsection{Applications}

The localization of BBHs, at least from GWs, is generally never truly accounted for. The localization of a black hole can break the degeneracy between the observed luminosity distance and the magnification as well as obtain an independent measurement of the Hubble constant \cite{Hannuksela2020}. 

The ability to identify the host of a BBH brings with it unique opportunities to study the relationship between mergers and their host galaxies, and the high-redshift population of galaxies producing lensed GWs. The ability to localize the host can tell us about the type of galaxies that produce BBH mergers at higher redshifts. Even with a single event, this will allow us to gain some information about the kind of galaxies hosting BBH, and with accumulating observations will allow for population studies to be conducted at a statistical level. 

Furthermore, Ref. \cite{Poon2024} showed that while a similarity transformation degeneracy \cite{Gorenstein1988, Saha2000} exists in GW lensing, EM information from the host can break that degeneracy. Ref. \cite{Poon2024} did, however, show that while the mass sheet degeneracy cannot be resolved from these observations alone even with the addition of EM information, obtaining a velocity dispersion measurement for the lens galaxy can help break that. This measurement is also only possible if the host and lens are identified, and therefore this localization is necessary in order to break the degeneracies present in both GW-only and multi-messenger lensing. 

Identification of the host-lens system with a GW counterpart also will allow us to probe the lens structure in great detail; effects such as millilensing from dark matter substructures can cause effects similar to flux-ratio anomalies in the GW amplitudes \cite{Mao1998} and affect the time delays \cite{Liao2018}.

In addition, as mentioned before, there is the risk that a host may not be identified. If this becomes a repeating pattern for lensed GWs, and we assume a case wherein we can safely presume a high lens catalogue completeness fraction, another avenue opens up: BBH kick velocities. There currently remain some open questions about the kicks with which BBH form \cite{Callister2021}. Repeated non-identification of the host lens could imply a similar case as has been observed for SNe, wherein the BBH may be kicked so far from its host that the host is not lensed. This avenue is as-yet unexplored, as it does not appear to be mentioned in literature.

This localization therefore has a variety of applications, and an even better understanding of the reaches of this method will come to light with the detection of real lensed GWs.

\section{Summary}\label{sec:conclusion}

We have summarized the current science towards localizing a lensed binary black hole, from the LVK detection up to the host identification, and asked the question: How much further can we probe this? A pin-point localization, at least to some theoretical approximations, is possible to the sub-arcsecond regime depending on the number of identified images. We present a simple Bayesian test of association between a possible host and a lensed GW. For a lensed event, previous studies found that it was possible to identify the host for \textit{Euclid-}observable lenses in about a third of cases, and upcoming telescopes, GW detectors, and statistical methods will improve this statistic. Following host localization, it is possible to then push this further to pin-point the binary black hole's position within its host to degenerate sub-arcsecond regions.

As a rather encouraging outlook to conclude this section and the limitations mentioned, with the advent of better and more instruments both in EM and GW, the exponential increase in the number of GWs identified, the ticking countdown towards detection of GW lensing, the number of catalogued lenses expected to rise by orders of magnitude, and the first lensed GW offering intriguing opportunity for immediate dedicated follow-up in any case, we expect to see these problems reduced to the point where it becomes plausible for us to identify the host of a lensed BBH, or any other dark lensed GW event. 

With it being a matter of \textit{when} rather than \textit{if}, the first lensed GW detection will prove an exciting opportunity for multi-messenger follow-up even for dark sirens without a direct EM counterpart from the source, opening up a new multi-messenger route to better understand BBHs and their hosts.

\ack{We thank the members of the LIGO-Virgo-KAGRA gravitational lensing group, as well as our collaborators from the EM side for a great number of useful discussions. We also thank the organizers of the Royal Society Multi-messenger Gravitational Lensing meeting for bringing this emergent community together, and its attendees for interesting discussions full of novel ideas and applications. JSCP and LEU are supported by the Hong Kong PhD Fellowship Scheme (HKPFS) from the Hong Kong Research Grants Council (RGC). LEU, HP, JSCP, and OAH acknowledge support by grants from the Research Grants Council of Hong Kong (Project No. CUHK 14304622 and 14307923), the start-up grant from the Chinese University of Hong Kong, and the Direct Grant for Research from the Research Committee of The Chinese University of Hong Kong. JJ is supported by the research program of the Netherlands Organisation for Scientific Research (NWO). The authors are grateful for computational resources provided by the LIGO Laboratory and supported by the National Science Foundation Grants No. PHY-0757058 and No. PHY-0823459. This material is based upon work supported by NSF’s LIGO Laboratory which is a major facility fully funded by the National Science Foundation.}


\newpage

\section*{Appendix A}

We make here explicit the details of the calculation to obtain the association evidence. 

\begin{align}
    \mathcal{Z_A} &= \int{p(d_{GW}, d_{EM} | \Vec{\theta}_{EM}) p(\Vec{\theta}_{EM}) d\Vec{\theta}_{EM}} \nonumber\\
    &= \int{ p(d_{GW}|d_{EM},\Vec{\theta}_{EM})p(d_{EM} | \Vec{\theta}_{EM}) p(\Vec{\theta}_{EM}) d\Vec{\theta}_{EM}} \nonumber\\
    &= \int{ p(d_{GW}|\Vec{\theta}_{EM}) p(d_{EM} | \Vec{\theta}_{EM}) p(\Vec{\theta}_{EM}) d\Vec{\theta}_{EM}} \nonumber\\
    &= \int{ p(d_{GW}|\Vec{\theta}_{EM}) p(d_{EM}|\Vec{\theta}_{EM}) \frac{p(\Vec{\theta}_{EM} | d_{EM})p(d_{EM})}{p(d_{EM}|\Vec{\theta}_{EM})}d\Vec{\theta}_{EM}} \nonumber\\ 
    &= \int{ p(d_{GW}|\Vec{\theta}_{EM}) p(\Vec{\theta}_{EM} | d_{EM})p(d_{EM}) d\Vec{\theta}_{EM}} \nonumber\\
    &= p(d_{EM}) \int{p(d_{GW} | \Vec{\theta}_{lens}, z_s, z_l, \beta}) p(\beta) p(z_s) p(z_l) p(\Vec{\theta}_{lens} | d_{EM})  d\Vec{\theta}_{EM}
\end{align}.

$\mathcal{Z_A}$ marks as before the evidence of association, $p(d_{EM})$ the evidence of the lens data which is the same regardless of the hypothesis and will be cancelled out in the Bayes' factor, $p(d_{GW} | \Vec{\theta}_{lens}, z_s, z_l, \beta)$ the likelihood we are interested in calculating, $p(\beta), p(z_s), p(z_l)$ the priors distributions of the source position, source redshift and lens redshift (respectively), and $p(\Vec{\theta}_{lens} | d_{EM})$ the posteriors obtained from \textsc{lenstronomy}.
\end{document}